# Pore network model of evaporation in porous media with continuous and discontinuous corner films

Rui Wu[a,b], Tao Zhang[a], Chao Ye[a], C.Y. Zhao[a,b],*, Evangelos Tsotsas[c], Abdolreza Kharaghani[c]

[a]School of Mechanical Engineering, Shanghai Jiao Tong University, Shanghai 200240, China.

[b]Key Laboratory for Power Machinery and Engineering, Ministry of Education, Shanghai Jiao Tong University, Shanghai 200240, China

[c]Chair of Thermal Process Engineering, Otto von Guericke University, P.O. 4120, 39106 Magdeburg, Germany

* Corresponding author Tel: +86 (0)21-34204541

E-mails:

ruiwu@sjtu.edu.cn (R. Wu)

changying.zhao@sjtu.edu.cn (C.Y. Zhao).

**Abstract**

During evaporation in porous media, two types of corner films are distinguished. A continuous corner film is connected to the bulk liquid, while a discontinuous one is not. To disclose their effects on evaporation in porous media, a pore network model with both continuous and discontinuous corner films is developed, which considers the capillary and viscous forces as well as the effects of corner films on the threshold pressures of pores. The capillary valve effect induced by the sudden geometrical expansion between the small and large pores is also taken into account in the model. The developed pore network model agrees well with the evaporation experiment with a quasi 2D micro model porous medium, in terms of not only the variation of the liquid saturation in each pore but also the variation of the total evaporation rate. The pore network models that neglect the corner films or the liquid viscosity are also compared with the experiment so as to shed light on the roles of the corner films. The continuous corner films, which contribute to sustain the high evaporation rate, can be interrupted to be the discontinuous ones not only by the gas invasion into pores but also by the capillary scissors effect due to the local convex topology of the solid matrix.

# 1. Introduction

During evaporation of a porous medium, the initially liquid-filled pores are gradually occupied by gas. This gas invasion determines the liquid distribution in the porous medium and hence the evaporation kinetics. In the course of gas invasion into pores in the porous medium, liquid can be retained in the corners of pores, forming the so-called corner films. Two types of corner films can be distinguished, as shown in Fig. 1. A continuous corner film is connected to a pore fully or partially saturated with liquid (liquid in such pore is called bulk liquid). A discontinuous corner film is not in contact with any bulk liquid. The effects of continuous and discontinuous corner films on the liquid transport are completely different. The continuous corner films provide hydraulic pathways for liquid transport from the interior of the porous medium towards the evaporative surface, sustaining the liquid saturation at the surface. This contributes to a high and relatively constant total evaporation rate, which forms the constant evaporation rate period. By contrast, the discontinuous corner films do not contribute to the liquid flow in the porous medium. Therefore, to unravel the mechanisms of evaporation in porous media, it is necessary to understand in detail the role of the continuous and discontinuous corner films.

Evaporation experiments with single capillary tubes have been performed so as to shed light on the roles of corner films [1 – 4]. The kinetics of evaporation in a square capillary tube with constant cross sectional area can be divided into three periods: a constant rate period, a falling rate period, and a receding front period, similar to that of porous media. The shape of the corner film at the entrance of a capillary tube is important to the evaporation rate. Nevertheless, the void space in a real porous medium is composed of hundreds and thousands of pores of various shapes and sizes. From this point of view, the evaporation of a single capillary tube with a constant cross sectional area may not fully reflect the transport phenomena in porous media.

Evaporation in porous media is a typical multi-scale process, which includes the interface scale, pore scale, and continuum scale. The interface scale processes, e.g., the capillary valve effect on the gas-liquid interface movement [5], can influence the transport processes at the continuum scale. Experimentally, it is difficult, if not possible, to observe directly the multi-scale transport processes in porous media, since real porous media are opaque and have complex microstructures, and the current state-of-the-art visualization techniques, e.g., X-ray or neutron imaging, cannot combine the visualization resolution and the size of the view field simultaneously. An alternative way is to opt for the pore network model (PNM).

The PNM, which has a good trade-off between the description of the interface and pore scale phenomena and the computational efficiency, is an ideal tool to reveal the multi-scale transport processes in porous media and hence has received more and more attention [6 – 11]. The basis of this model is to approximate the void space of a porous material by a pore network composed of regular pores, i.e. large pore bodies connected by small pore throats (e.g., Fig. 1). Liquid and vapor transport in each pore are then described, from which the evaporation process can be unraveled. Two types of PNMs can be discerned. In the dynamic PNMs (DPNMs), the liquid viscosity is considered; whereas it is neglected in the quasi-static PNMs (QSPNMs).

Developing an accurate PNM is the cornerstone to employ it for understanding of the multi-scale transport processes in porous media. Up to now, the corner films in PNMs, if considered, are commonly assumed as continuous ones [12 – 16]. Vorhauer et al. [17] developed a 2D QSPNM that accounts for both the discontinuous corner films, and revealed that the PNM with both continuous and discontinuous corner films leads to better predictions on the distribution of corner films in the pore network than the PNM with only continuous corner films. Hence, the discontinuous corner films must be considered in PNM so as to understand in detail the evaporation process in porous media. Nevertheless, the liquid viscous forces are not considered in the QSPNM of Vorhauer et al. [17].

For evaporation in porous media with corner films, the liquid flow resistances in corner films are obviously larger than those in pores. The liquid flow along the corner films hence may not compensate the liquid loss at the evaporative surface. But in the QSPNM, the evaporation loss is assumed to be completely compensated by the capillary liquid flow. Furthermore, in order to investigate the transport of solute, e.g., salt [18, 19], in pore network during evaporation, the liquid velocity filed is needed. Therefore, developing a DPNM that considers the liquid viscous forces as well as the continuous and discontinuous corner films is of critical importance to explore the evaporation processes in porous media.

In this work, a DPNM of evaporation in pore networks with both the continuous and discontinuous corner films is successfully developed, which considers capillary and liquid viscous forces, the continuous and discontinuous corner films, the effect of the corner films on the threshold pressures of pores, and the capillary valve effect induced by the sudden geometrical expansion at the interface between the small and large pores [20, 21]. The developed DPNM predicts well the variation of the liquid distribution and the total evaporation rate in the evaporation experiment with a quasi 2D microfluidic pore network, validating the effectiveness of the developed DPNM, which paves a way to understand the multi-scale evaporation processes in porous media with both continuous and discontinuous corner films.

In what follows, the evaporation experiment with the quasi 2D microfluidic pore network is introduced. The developed DPNM is depicted in detail in Section 3. The main results are presented in Section 4. In Section 5, interruption of the corner films due to the capillary scissor effect is discussed. Finally, conclusions are drawn in Section 6.

## 2. Experiment

The microfluidic pore network used for the evaporation experiment is fabricated by bonding a silicon wafer etched with the designed pore structures to a glass sheet. The pore network consists of 5×5 large pore bodies connected by small pore throats, see Fig. 2a. In order to validate the PNM developed in section 3 conveniently, the pore network is composed regular cuboid pores, since the liquid and gas transport in cuboid pores can be easily described. Large pore bodies connected by small pore throats are used in the pore network to elucidate the capillary valve effect induced by sudden geometrical expansion.

All the pores in the pore network have the same height of 50 μm in the direction perpendicular to the plane shown in Fig. 2a. Hence, we concern pore and meniscus structures mainly in this plane hereafter unless otherwise specified. All the pore bodies are square and have a side length of $l = 1$ mm. The distance between the centers of two adjacent pore bodies is 2 mm. All the pore throats are rectangular and have the same length of $l = 1$ mm. The widths of pore throats, $w$, are randomly distributed in the range of 0.14 – 0.94 mm. The pore network is connected to the environment through a pore throat of 2 mm long and 0.5 mm wide. This pore throat is also called the outlet pore. The width of each pore throat can be found in Fig. S1 in the supporting materials.

The cross section of a pore in the pore network is rectangular, Fig. 2b. However, the corner is actually rounded, not sharp (more details can be found in Section 5). A pore has four walls: a top glass wall, two side silicon walls, and one bottom silicon wall. Each solid element, surrounded by pores, has eight side walls, Fig. 2a. The length of each side wall of a solid element can be determined by the sizes of connected pores. Note that a side wall can be connected to two corner films (Fig. 2c): one is at the corner between the side and top walls, and the other is at the corner between the side and bottom walls. A side wall is saturated or filled when it is covered by corner films. The contact angle is defined as the angle between the gas-liquid (ethanol) interface and the liquid-solid interface. The contact angle for the silicon wall, $\theta_{si} = 35^\circ$, is evaluated based on the shape of gas-liquid interface (meniscus) obtained in the evaporation experiment. The contact angle for the glass wall is $\theta_g = 0^o$ [22].

The pore network is initially immersed into ethanol (purity 99.99%) contained in a plastic cylinder depressurized by a vacuum pump so as to saturate the pore network with ethanol. Then the pore network is horizontally placed in a chamber with almost constant temperature ($25 \pm 2°C$) and relative humidity ($32 \pm 2\%$). Time variation of the liquid distribution in the pore network is recorded by a camera (Nikon D810) equipped with a macro lens (AF-S VR Micro-Nikkor 105 mm f/2.8 IF-ED).

Since the cross section of the pore in the pore network is rectangular, it is not easy to observe the corner films directly in the pore network during evaporation. But the variation of the liquid configuration in the pore network can indicate the existence of corner films. As shown in Fig. 3, during evaporation, gas invades into pore body A, transforming the meniscus curvature from concave to convex (to the gas phase), see the images from $t = 0$ to 70 min. This is probably due to the contaminants in the pore body A. As a result, the liquid pressure increases in the pore body A, thereby resulting in higher liquid pressure in other pores, e.g., pore body B. The increase of liquid pressure leads the corner films in the pore body B, if any, to swell. This swelling of corner films explains well why the pore body B is filled by liquid during the gas invasion into pore body A, see image of $t = 70$ min in Fig. 3. As the meniscus in pore body A becomes concave again, the pressure of liquid is decreased, resulting in the shrinkage of liquid in the pore body B, see image of $t = 120$ min in Fig. 3. The variation of the liquid configuration in pore body B demonstrates that there are continuous corner films connecting the pore body B and the bulk liquid in the pore network shown in Fig. 3.

Moreover, menisci are white and bright in the experimental images, Fig. 3. Some walls attached to empty pores (e.g., pore body B) show similar color as menisci, whereas some others (e.g., the walls of solid element C) do not. By comparing the colors of walls and menisci, we can also infer that there exist corner films in the pore network.

## 3. Pore network model

A 2D dynamic pore network model (DPNM) with continuous and discontinuous corner films is developed. The developed model considers the capillary forces, the liquid viscous forces, the continuous and discontinuous corner films, the effects of the corner films on the threshold pressure of pores, the capillary valve effect induced by the sudden geometrical expansion at the interface between small and large pores, and the vapor transport outside of the pore network. Since the aim of the present study is to develop and validate a new DPNM with continuous and discontinuous corner films, only the evaporation processes in the 2D pore network used in the experiment are modeled. The 3D pore network simulations will be studied in a future work.

To validate the developed DPNM, the pore network used in the model mimics the pore structure shown in Fig. 2a. To calculate the evaporation rate from the pore network, a cubic boundary zone is introduced right above the pore network surface. The ambient vapor pressure ($P_{v,e} = 0$) is applied to all sides of the boundary zone except the side adjacent to the pore network. At the side adjacent to the pore network, the area attached to the outlet pore of the pore network is taken as an inner surface, while the other area is non-permeable. To evaluate the size of the boundary zone, the vapor pressure at the inner surface between the pore network and the boundary zone is taken as the saturated vapor pressure ($P_{v,s}$ = 9.171 ×10$^3$ Pa), and the evaporation rate from this inner surface is calculated. The vapor transport is considered as the steady vapor diffusion through the stagnant air, and can be described by [23]:

$$\nabla \cdot \left( \frac{DP_g}{RT} \frac{1}{P_g - P_v} \nabla P_v \right) = 0 \tag{1}$$

where $P_g$ is the total gas pressure (1.013 ×10$^5$ Pa), $D$ the vapor diffusivity (1.264 ×10$^{-5}$ m$^2$/s), $R$ the universal gas constant (8.314 J/(mol K)), $T$ the ambient temperature (298 K).

The finite volume method [24] is used to solve Eq. (1) in the boundary zone. First, the boundary zone is divided into a number of cubic grid cells, each of which has the same side length of 50 μm. Then, Eq. (1) is discretized based on the central difference scheme. The variables are stored at the center of each grid cell. Hence, a set of linear equations for the vapor pressure field is generated and solved by the BiCGSTAB method [25]. The evaporation rate is determined for various boundary zone sizes. The results indicate that the cubic boundary zone with the side length of 5 mm is enough for determination of the evaporation rate.

In the present DPNM model, we assume that the vapor pressure at a filled side wall is equal to the saturated vapor pressure. In previous studies, 1D + 1D model has been used for vapor transport in a pore with corner film, e.g., [13]. In this 1D + 1D model, the 1D vapor transfer from the corner film to the center of the pore is considered as the source term in the 1D vapor transport along the centerline of the pore. However, we find that the 1D + 1D model cannot capture the vapor transport in the pores with corner films in the present study, Fig. S2 in the supporting materials. Hence, the 2D model based on Eq. (1) is used to describe the vapor transport in the pore network, which, obviously, increases the computational complexity and costs.

We find that if the side walls of the outlet pore and the side walls at edges of the pore network are filled, the vapor inside the pore network is saturated even though the vapor pressure at the entrance of the outlet pore is zero, Fig. S3 in the supporting materials. Since the size of the pore network is not large, our theoretical analysis (see following) indicates that there will be continuous corner films at edges of the pore network and the outlet pore even though only one pore at edges of the pore network is filled. Hence, we assume that the side walls of the outlet pore are filled, and that the vapor inside the pore network is saturated. To this end, the evaporation rate from the pore network can be determined by solving Eq. (1) in the 3D boundary zone and in the 2D outlet pore by using the finite volume method mentioned above. The outlet pore is divided into a number of square grid cells with the same side length of 50 μm. A grid cell at the entrance of the outlet pore is adjacent to only one grid cell in the boundary zone.

Liquid flow in the pore network is assumed as a fully developed laminar flow. The liquid pressure in a pore, $P_l$, is calculated at the pore center. The mass flow rate between two adjacent liquid filled pores, e.g., from pore body $pj$ to pore throat $pi$, is determined as [26]:

$$Q_{f,pj \to pi} = \frac{16 w_{pi} \rho_l h^3}{\pi^4 \mu_l} \frac{P_{l,pj} - P_{l,pi}}{l_{pj} + l_{pi}} \sum_{n=0}^{\infty} \frac{1}{(1+2n)^4} \left(1 - \frac{\tanh[(1+2n)\pi w_{pi}/2h]}{(1+2n)\pi w_{pi}/2h}\right) \quad (2)$$

where $\rho_l$ and $\mu_l$ are the density and viscosity of liquid, respectively. As reported in [26] if $n = 0$ and 1 is used in Eq. (2), the calculation error is less than 1% when the aspect ratio, $h/w_{pi}$, is less than 2. For the pore network in the present study, the aspect ratio is smaller than 0.2. Hence, only $n = 0$ and 1 are used in Eq. (2).

Each solid element inside the pore network has eight side walls, and a side wall is also the interface between a solid element and a pore, Fig. 2a. A side wall can be connected to two corner films, Fig. 2c. We assume that the liquid pressure files along the corner films attached to the same side wall are the same. The curvature of corner film in the plane parallel to the liquid flow direction is neglected. As in previous studies, e.g., [13], the mass flow rate along the flow direction of a corner film (e.g., $x$ direction) is determined as:

$$Q_c = -\rho_l \frac{\alpha r_c^4}{\mu_l} \frac{dP_{l,c}}{dx} \quad (3)$$

where $r_c$ is the radius of curvature of the corner film in the plane perpendicular to the liquid flow direction, and its value depends on the pressure of the liquid in the corner film:

$$r_c = \frac{\sigma}{P_g - P_{l,c}} \quad (4)$$

where $\sigma = 0.022$ N/m is the surface tension.

In Eq. (3), $\alpha$ is a dimensionless parameter, which is defined as the ratio of the shape factor to the dimensionless resistance [27]. Based on the work of [27], it can be concluded that the dimensionless parameter $\alpha$ depends on the geometry of the corner and the contact angle. To determine the value of $\alpha$, we compute the distribution of liquid velocity in the plane perpendicular to the corner flow direction. The inertial forces are neglected for the liquid flow in the corner films because of low flow velocity. The liquid velocity distribution is computed by COMSOL, from which the mass flow rate is obtained. Then, the value of $\alpha$ is determined based on Eq. (3). The detailed calculation can be found in Appendix A. The dimensionless parameter $\alpha$ is $\alpha_t = 7.02 \times 10^{-4}$ for the corner film attached to the top wall, and $\alpha_b = 9.25 \times 10^{-5}$ for that connected to the bottom wall.

The liquid pressure in the corner films attached to a side wall is determined at the middle point of this side wall in liquid flow direction. The curvature radius of the corner film at a side wall in the plane perpendicular to the flow direction is constant. The mass flow rate from corner films at side wall $sj$ to those at the adjacent side wall $si$ connected to an empty pore is thus determined as:

$$Q_{c,sj \to si} = \frac{2}{\mu_l} \frac{(\alpha_{t,si} + \alpha_{b,si})(\alpha_{t,sj} + \alpha_{b,sj}) r_{c,si}^4 r_{c,sj}^4}{l_{c,si}(\alpha_{t,si} + \alpha_{b,si}) r_{c,si}^4 + l_{c,sj}(\alpha_{t,sj} + \alpha_{b,sj}) r_{c,sj}^4} (P_{l,sj} - P_{l,si}) \qquad (5)$$

where $l_c$ is the length of the corner film, equal to the length of the attached side wall. Here, we consider the liquid flow between two corner films, at least one of which is in an empty pore with no bulk liquid. A filled pore has bulk liquid. The liquid flow between two corner films in filled pores is included in the flow between two adjacent pores, Eq. (2). The liquid pressure of the corner films in a filled pore is equal to that of this pore.

During evaporation, the liquid in the pore network is gradually replaced by the gas phase, resembling gas invasion into a liquid filled pore network. This gas invasion process depends on the threshold pressure of each pore and corner. A meniscus can invade a pore (or a corner) only when the pressure difference, $P_g - P_l$, across this meniscus is larger than the threshold pressure of the pore (or the corner). Because of the capillary valve effect, which is induced by the sudden geometrical expansion at the interface between a pore throat and a pore body, the threshold pressure for gas invasion into a filled pore body $pi$ from the adjacent pore throat $pj$ is determined as:

$$P_{t,pi} = \sigma \left( \frac{2\sin[max(90°, \theta_{silicon})]}{w_{pj}} + \frac{1}{r_h} \right) \tag{6}$$

where $r_h$ is the curvature radius of the meniscus in the plane parallel to the height direction, which is the same for all pores:

$$r_h = \frac{h}{1 + cos\theta_{si}} \tag{7}$$

The threshold pressure for gas to invade a filled pore throat $pi$ from the adjacent pore body is gained based on the MSP method [28, 29]:

$$P_{t,pi} = -\sigma \frac{k_1 + k_2 r_c^*}{w_{pi}h - k_3 r_c^{*2}} \tag{8}$$

The derivation of Eq. (8) and the values of $k_1$, $k_2$, $k_3$ and $r_c^*$ are presented in the Appendix B. The threshold pressure for gas to invade a corner film attached to side wall $si$ is determined as:

$$P_{t,si} = \frac{\sigma}{k_c} \tag{9}$$

where $k_c$ is the critical radius. We find that the influence of $k_c$ is trivial when $k_c$ is small, e.g., < 5 μm. Here, $k_c = 1$ μm is used.

Both the continuous and discontinuous corner films are considered in the present

PNM. The continuous corner films provide liquid flow paths between filled pores. Hence, it is needed to take into account the corner films in order to identify the liquid cluster in the pore network. The procedures for liquid cluster identification are as follows:

(1) Each filled pore and filled solid element have a cluster label, CL, and a cluster number, CN. Initially, CN = 0, and CL = 0 is applied to all the filled pores and the filled solid elements.

(2) Each filled pore body is scanned. If a filled pore body has a cluster label of CL = 0, then CN = CN + 1, and CL = CN is applied to this pore body. Then, we scan each filled pore and each filled solid element; if one has CL = CN, we set the adjacent filled pores and filled solid elements with CL = 0 to have CL = CN; this identification is repeated until no filled pores or filled solid elements with CL = CN have connected filled pores or filled solid element with CL = 0.

(3) Each filled pore throat is scanned. If a filled pore throat has a cluster label of CL = 0, then CN = CN + 1, and CL = CN is applied to this pore throat. Then, identification process similar to that in step (2) is performed to determine the liquid cluster composed of filled pore throats and filled solid elements with CL = CN. We do not consider the filled pore bodies in this identification process. The reason is that all the pore bodies have been identified or labeled in step (2); hence, in this step, we identify the liquid cluster with pore throats but without pore bodies.

(4) Each filled solid element is scanned. Since all the filled pores have been scanned in steps (2) and (3), if there is a filled solid element with CL = 0, it must be the isolated one and has no connected filled pores. Hence, if a filled solid element has a cluster label of CL = 0, then CN = CN + 1, and CL = CN is applied to this solid element, which is also considered as a liquid cluster. All menisci attached to this isolated filled solid element are set to be moving meniscus.

The following algorithm is used to simulate the evaporation process in the pore network.

(1) Initially, the pore network is fully saturated with liquid except the outlet pore, which is empty but has liquid in its four corners, i.e., corner films. All the menisci are static.

(2) The vapor pressure field in the pore network is determined. The vapor pressure inside the pore network is at the saturated vapor pressure; whereas the vapor pressure in the outlet pore is determined by solving Eq. (1).

(3) The evaporation rate from the pore network is calculated. Since the vapor inside the pore network is saturated, the evaporation rate is zero from the saturated side walls and pores inside the pore network. The evaporation rate from a side wall *si* in the outlet pore is:

$$Q_{d,si} = \sum_{i=1}^{m} 2Dh \frac{M_v P_g}{RT} \ln \frac{P_g - P_{v,c,i}}{P_g - P_{v,s}} \tag{10}$$

where $P_{v,c}$ is the vapor pressure in the grid cell, and *m* is the number of grid cells along the length direction of the outlet pore. The evaporation rate from the pore network is equal to the sum of the evaporation rates from the two side walls in the outlet pore.

(4) The liquid clusters in the pore network are identified. Then, the states (moving or

static) of menisci in each liquid cluster are checked. If all the menisci in a liquid cluster are static, the meniscus in the filled pore with the lowest threshold is set to be moving. The corners have a larger threshold pressure than pores.

(5) The liquid pressure in filled pores and filled side walls are determined based on the mass conservation law, e.g., for a filled pore *pi* or a filled side wall *si*, we have:

$$\sum_{pj} Q_{f,pj \to p(s)i} + \sum_{sj} Q_{f,sj \to p(s)i} + \sum_{pk} Q_{d,pk \to p(s)i} = 0 \tag{11}$$

where $Q_{f,pj \to p(s)i}$, $Q_{f,sj \to p(s)i}$, and $Q_{d,pk \to p(s)i}$ are the rates of mass transport into the filled pore *pi* or the filled side wall *si* through the flow from the adjacent filled pore *pj*, through the flow from the adjacent filled side wall *sj*, and through the diffusion from the adjacent empty pore *pk*, respectively. Note that the mass diffusion rate inside the pore network is zero because of the saturated vapor.

(6) For each liquid cluster, the pressure difference $P_d$ is determined for each filled pore *pi* with static menisci:

$$P_d = P_g - P_{l,pi} - P_{t,pi} \tag{12}$$

Then, the filled pores with menisci and positive $P_d$ are scanned, among which the static menisci in the pore with the largest $P_d$ are set to be moving.

(7) Repeat steps (5) and (6) until there is no more change of the meniscus state in each liquid cluster.

(8) For each filled pore *pi* or filled solid element *ei* with moving menisci, the liquid removal rate is determined as:

$$Q_{out,p(e)i} = \sum_{pj} Q_{f,pi \to pj} + \sum_{sj} Q_{f,pi \to sj} + \sum_{pk} Q_{d,p(e)i \to pk} \tag{13}$$

The definition of $Q_{f,\,pi \to pj}$, $Q_{f,\,pi \to sj}$ and $Q_{d,\,pi \to pk}$ can be found in Eq. (11). $Q_{d,\,ei \to pk}$ is the mass diffusion rate from solid element *ei* to adjacent empty pore *pk*, which equals to the evaporation rate from the side wall between solid element *ei* and pore *pk*. It should be noted a filled solid element has moving menisci when it is isolated, and the liquid surrounded such solid element can be removed only through the vapor diffusion. Hence, the liquid flow out of the solid element is not considered in Eq. (13).

(9) For each filled pore *pi* or each filled solid element *ei* with moving menisci, the time to empty the liquid therein is determined:

$$t_{p(e)i} = \frac{\rho_l V_{l,p(e)i}}{Q_{out,p(e)i}} \tag{14}$$

The minimum time $t_{min}$ is selected as the step time. In Eq. (14), $V_{l,pi}$ is the volume of the bulk liquid in pore *pi*, and $V_{l,ei}$ is the volume of corner films attached to solid element *ei*.

The volume of corner films attached to a solid element *ei* with static menisci (i.e., the solid element is connected to at least one filled pore) is the sum over corner film volumes at the eight side walls of this solid element:

$$\begin{aligned}V_{l,ei} = &\sum_{si} \left( r_{c,si}^2 \cos\theta_{si} - \frac{r_{c,si}^2 \cos\theta_{si}\sin\theta_{si}}{2} - \frac{\pi - 2\theta_{si}}{4\pi}\pi r_{c,si}^2 \right) l_{si} \\ &+ \sum_{si} \left( r_{c,si}^2 \cos^2\theta_{si} - r_{c,si}^2 \cos\theta_{si}\sin\theta_{si} - \frac{\pi - 4\theta_{si}}{4\pi}\pi r_{c,si}^2 \right) l_{si}\end{aligned} \tag{15}$$

where $r_c$, determined by Eq. (4), is dependent on the pressure of liquid in the corner film. The first and the second term at the right hand of Eq. (15) are the volumes of the corner films (at the side wall *si*) attached to the top and bottom walls, respectively. If the filled solid element *ei* has moving menisci, then the volume of the corner films can be changed during evaporation, and is determined by following Eq. (17).

The volume of bulk liquid in a filled pore *pi* with static menisci and liquid saturation of one is equal to the difference between the volume of liquid in the pore and the volume of corner films in the pore:

$$V_{l,pi} = V_{pi} - \sum_{si} V_{l,si} \tag{16}$$

The first term at the right hand of Eq. (16) is the volume of pore *pi*, and the second term is the sum of the volumes of corner films in the pore. The volume of bulk liquid in a filled pore with moving menisci is determined by the following Eq. (17).

(9) The volume of liquid in each filled pore *pi* or filled solid element *ei* with moving menisci is updated:

$$V_{l,p(e)i} = V_{l,p(e)i} - \frac{t_{min}Q_{out,p(e)i}}{\rho_l} \tag{17}$$

(10) Repeat from step (2) until the prescribed condition is reached.

For better understanding of the effects of the corner films, the evaporation in the pore network is simulated by not only the DPNM with corner films (DPNM_wC) introduced above but also the dynamic pore network model without corner films (DPNM_woC). The detailed algorithm of the DPNM_woC can be found in [5, 30]. In addition, the quasi-static PNM with corner films (QSPNM_wC) is also used so as to elucidate the role of liquid viscous forces. In the QSPNM_wC, the liquid viscous forces are not considered, and for each liquid cluster, only the meniscus in the filled pore or filled solid element with the lowest threshold pressure is set to be moving.

## 4. Results

To reveal the role of corner films in evaporation of porous media, the simulation results obtained from various PNMs are compared with the experimental data. The evolution of liquid distribution in the pore network during evaporation is shown in Fig. 4. In the simulation results, the gas, liquid, solid are shown in white, red, and gray, respectively. In the results obtained from the DPNM_wC, the corner films, attached to solid elements, are also shown in red. From the beginning of the evaporation process until the total liquid saturation reaches $S_t \geqslant 0.52$, the results obtained from the DPNM_wC and the QSPNM_wC are almost identical. Hence, in Figs. 4 – 7, the results of the QSPNM_wC are not shown. The difference between the DPNM_wC and the QSPNM_wC will be discussed in detail later.

In the experimental results, the gas, liquid, solid are shown in light gray, light black, and dark gray, respectively. It is not easy to observe the corner films directly in the experimental images, although the gas-liquid interfaces are white and bright. In the experiment, two small gas bubbles exist at the bottom-left and bottom-right of the pore network. These two gas bubbles remain static during evaporation, Fig. 4; hence, their influences are negligible.

We do not compare the experimental and simulation results when the total liquid saturation is $S_t < 0.52$. The liquid distribution at $S_t = 0.52$ is shown in Fig. 4. After this moment, gas invades the second pore body from the left at the bottom of the pore network. By embarking this invasion, the meniscus in the pore body becomes unstable, and enters the pores at the lower-left zone of the pore network quickly; at the same time, some empty pores at the center of the pore network are refilled by liquid, Fig.

S4 in the supporting materials. This unstable phenomenon is not considered in the present PNM. Hence, we compare simulations and measurements mainly at $S_t \geqslant 0.52$.

During evaporation, liquid is gradually replaced by gas in the pore network, and large liquid clusters are split into smaller ones. At $S_t = 0.82$, an isolated filled solid element is observed in the DPNM_wC; the corner films attached to this solid element are discontinuous, since they are not connected to any bulk liquid, Fig. 4. Since the gas phase is fully saturated with vapor, these discontinuous corner films always exist during evaporation in the DPNM_wC. Experimentally, we also observe that this solid element is covered by liquid, Fig. S5 in the supporting materials. However, in the DPNM_woC, corner films are not considered. Hence, the number of liquid clusters, as shown in Fig. 5, is $N_t = 2$ in the DPNM_wC, whereas $N_t = 1$ in the DPNM_woC at $S_t = 0.82$.

In the DPNM_woC, two liquid clusters are discerned when gas invades the pore body at the middle of the right edge of the pore network, image of $S_t = 0.75$ in Fig. 4. The liquid cluster at the upper-right zone of the pore network is smaller. The evaporation rates from the large and the small liquid clusters are $6.67 \times 10^{-16}$ and $3.15 \times 10^{-17}$ kg/s, respectively. By contrast, In the DPNM_wC, when gas invades the pore body at the middle of the right edge, the number of liquid clusters remains unaltered (Fig. 5), since the continuous corner films connect the filled pores.

As evaporation proceeds (from $S_t = 0.75$ to 0.58), the two pore bodies at the upper-right zone are emptied in the DPNM_wC, consistent with the experimental observations. But in the DPNM_woC, only one pore body is emptied. To understand the difference between these two PNMs, we present in Fig. 6 the detailed gas invasion process in the upper-right zone of the pore network. In the DPNM_wC, after gas invasion into pore body C (stage II in Fig. 6), pore throats A and D are still in the same liquid cluster, because they are connected via the continuous corner films. Pore throat D, 520 μm wide, has a lower threshold pressure than pore throat A of 340 μm wide. Hence, pore throat D is invaded in stage III, followed by pore body E in stage IV.

By contrast, after gas invasion into pore body C in the DPNM_woC, filled pore throat A becomes the isolated liquid cluster, stage II in Fig. 6. Since the volume of liquid in pore throat A is rather small, it is emptied in stage III. Then pore throat D is invaded in stage IV. Evolution of liquid distribution in stages II and III also explains why the number of liquid clusters fluctuates during evaporation in the DPNM_woC, Fig. 5.

As evaporation proceeds (from $S_t = 0.58$ to $0.52$), gas invades the lower-left zone of the pore network. Both the experiment and the DPNM_wC reveal a filled pore throat between two empty pore bodies, image of $S_t = 0.52$ in Fig. 4. However, such filled pore throat is not observed in the DPNM_woC. Moreover, as evaporation continues to $S_t = 0.49$, the DPNM_woC shows that the liquid at the upper-right zone is completely removed; whereas both the experiment and the DPNM_wC reveal that the liquid filled pores in this zone are not invaded by gas, since they are smaller and have higher threshold pressure than pores in lower-left zone, Fig. S6 in the supporting materials. Note that owing to the above mentioned unstable phenomenon in the experiment, the agreement is not good in the liquid distribution at the lower-left zone of the pore network between the experiment and the DPNM_wC at $S_t = 0.49$.

As evaporation continues further, both the experiment and the DPNM_wC reveal that filled pore throat A at the upper-right zone of the pore network becomes empty, whereas, in the QSPNM_wC, for which viscous forces are not considered, pore throat A is still filled, Fig. S7 in the supporting materials. As gas invades the lower-left zone (as depicted in Fig. 4), the liquid flow resistance from this zone to the filled pores at the upper-right zone increases, since the length of the corner films increases. To supply the liquid from the lower-left zone to the outlet pore so as to replenish the evaporation loss (which is constant), the liquid pressure in the filled pores at the upper-right zone must decrease (since the liquid flow resistance increases). The filled pore at the upper-right zone is invaded by gas when the liquid pressure in this pore decreases to the value that the pressure difference across the meniscus in this pore, $P_g - P_l$, is equal to or larger than its threshold pressure. Consequently, for evaporation in the pore networks with corner films, the liquid flow induced pressure loss must be taken into account, even for the pore network with small size, e.g., the one in the present study.

As shown in Fig. 4, evolution of the liquid distribution in the pore network during evaporation can be well predicted by the DPNM_wC, validating its effectiveness. The corner films at the edges of the pore network play an important role in connecting the liquid filled zones in the pore network. The solid elements at the edge of pore network are connected to each other. Hence, corner films at edges of the pore network can be easily connected to bulk liquid if there are filled pores at edges. In the DPNM_wC, if only one pore at the edges of the pore network is filled, then the corner films at the edge solid elements are the continuous ones and provide the flow paths between the filled zone inside the pore network and the corner films at the outlet pore. This is why corner films always exist in the outlet pore during evaporation in the DPNM_wC, Fig. 4.

In the DPNM_wC, the evaporation rate from the partially filled pores inside the pore network is zero since the gas phase is fully saturated with vapor in the presence of corner films at the outlet pore. To this end, the generated liquid clusters that are not connected to the outlet pore will not be emptied. As a result, the number of liquid clusters, as shown in Fig. 5, is always increasing during evaporation. In the DPNM_woC, however, the corner films are not taken into account. The gas phase inside the pore network is not saturated with vapor and this leads to non-zero evaporation rates from the filled pores. As a result, the liquid clusters are generated and removed during evaporation; and hence the number of liquid clusters is not monotonically increasing, Fig. 5.

The evolution of the total liquid saturation with the evaporation time obtained by the experiment and PNMs are compared in Fig. 7. For the experimental results, the total liquid saturation is obtained based on the image analysis. We also conduct an experiment to measure the weight change of the microfluidic pore network by using an electronic balance (Sartorius BT25S, Germany) so as to gain variation of the total liquid saturation during evaporation. The results obtained by these two methods are similar, Fig. S8 in the supporting materials. Since the corner films cannot be discerned in the image analysis, the good agreement between these two methods indicates that the volume of corner films is not significant (but the corner films play an important role in the evaporation processes, as discussed above).

As shown in Fig. 7, the evaporation process predicted by the DPNM_woC is slower than the experimental result and the one predicted by the DPNM_wC. This is owing to the fact that the corner films are not considered in the DPNM_woC. Not surprisingly, the liquid saturation varies linearly with the evaporation time in the experiment and the DPNM_wC, since the corner films at the outlet pore results in a constant evaporation rate from the pore network as $S_t$ decreases from 1 to 0.52. Interestingly, this linear variation is also observed in DPNM_woC. The reason is that the small pore throats near the outlet pore (Fig. 4) are filled with liquid during evaporation. As a result, the pore body connected to the outlet pore has almost constant vapor pressure, which in turn leads to an almost constant evaporation rate from the pore network.

As shown in Fig. 4, the evaporation induced gas invasion in the pore network is a random process. The gas invasion determines the liquid velocity in the pore network. The random gas invasion can result in oscillation of the liquid velocity. To reveal the influence of gas invasion on the liquid velocity field, we present in Fig. 8 the variation of the averaged liquid velocity, predicted by the DPNM_wC, in the pore throat B shown in Fig. 4. The flow from the left to the right is defined as the positive direction. The liquid velocity fluctuates during evaporation. Because of this fluctuation, it seems that the time averaging method, which has been used in the turbulent flow in porous media, e.g., [31], needs to be considered to combine with the volume-averaging technique [32] or averaging-thermodynamic method [33-36] so as to develop the continuum model for evaporation (also two-phase transport) in porous media. The role of the velocity fluctuation in two-phase transport in porous media will be explored in detail in a future study.

## 5. Discussion

The evaporation rate curve deduced from experiments, as shown in Fig. S8, can be divided into two stages: a relative high evaporation rate stage at $S_t > 0.2$, and a lower evaporation rate stage at $S_t < 0.2$. Since the simulations based on the experiment and the DPNM_wC are in good agreement (Figs. 4 and 7), and the DPNM_wC predicts corner films at the outlet pore, we believe that corner films also exist at the outlet pore in the experiment when $S_t > 0.2$. The reduction of the evaporation rate at $S_t < 0.2$ indicates that the corner films at the outlet pore are removed, although at $S_t = 0.1$ there are still filled pores at the edges of pore network. However, the DPNM_wC predicts that corner films exist in the outlet pore even if only one pore at edges of the pore network is filled. The discrepancy in the term of the evaporation rate between the DPNM_wC and the experiment at $S_t < 0.2$ implies that some factors are not considered in the model.

To understand the change of the evaporation rate observed in the experiment, we present in Fig. 9 the liquid distribution in the zone near the outlet pore from the evaporation time $t$ = 322.7 to 351.2 min. These two evaporation times correspond to the total liquid saturation $S_t$ = 0.25 and 0.18, which are in the transition region between the high and low evaporation rate stages. Evolution of the liquid distribution in the pore network during the evaporation process from $t$ = 322.7 to 351.2 min is presented in Fig. S9 in the supporting materials.

As shown in Fig. 9, as gas invades pore body F, residual liquid can be clearly observed in pore body I adjacent to the outlet pore. The side walls in the outlet pore are white and bright, indicating existence of corner films. As gas invades pore throat G, pore body I is connected to liquid in pore throat G through corner films, increasing the resistance for liquid flow from moving menisci to corner films in the outlet pore. As a result, liquid pressures in the corner films in the outlet pore and in pore boy I are reduced, and residual liquid in pore body I is rather small (image of $t$ = 344.1 min in Fig. 9).

After pore throat G is emptied (image of $t$ = 351.2 min in Fig. 9), the resistance is rather large for liquid flow from the moving menisci at the lower-right zone of the pore network to the corner films at the outlet pore, Fig. S9 in the supporting materials. Hence, liquid pressure in the corner films in outlet pore and in filled pore throat H is rather small so as to pump enough liquid to compensate the evaporative loss at the outlet pore. Since pores have smaller threshold pressures than corners, the DPNM_wC predicts that pore throat H will be invaded before the corner films in the outlet pore. In the experiment, however, pore throat H is still filled; whereas the change of color and brightness of the side wall in the outlet pore indicates that the corner films therein are removed, see the right two images in Fig. 9. We believe that the corner films removal in the outlet pore is the main reason for the two evaporation stages observed in the experiment, Fig. S8 in the supporting materials.

We find that removal of the corner films in the outlet pore is attributed to an effect referred to as the "capillary scissors". This effect is induced by the local convex topology of the solid matrix. We show in Fig. 10 schematically the corner films at intersection between side wall 1 in a pore throat and side wall 2 in a pore body. The intersection of these two side walls is not sharp but rounded. Such roundness also exists at the intersection between the side wall and the top (or bottom) wall, see Fig. S10 in the supporting materials. In the B – B plane shown in Fig. 10, the solid element is convex towards the fluid at the intersection of side walls 1 and 2. The pressure difference across the meniscus at this intersection is $P_g - P_l = \sigma(1/r_h - 1/r_i)$. Here, $r_h$ and $r_i$ are the curvature radii shown in Fig. 10, both of which are positive.

If we assume that liquid pressure in the corner film at the intersection point is the same as that in the neighboring side walls, e.g., side walls 1 and 2 in Fig. 10, then $r_h$ at the intersection point is smaller, since $1/r_i$ is zero for menisci at side walls. As the liquid pressure decreases, $r_h$ reduces. When $r_h$ reduces to the curvature radius, $r_{c,h}$, of the intersection between the side wall and the top (or bottom) wall, the corner films at side walls 1 and 2 are separated; at this moment, $r_i$ is equal to the curvature radius of the intersection between side walls 1 and 2, $r_{c,i}$. Hence, the threshold pressure for gas to invade the corner film at the convex intersection between two side walls is $\sigma(1/r_{c,h} - 1/r_{c,i})$, which decreases with smaller $r_{c,i}$, and hence could be lower than that of a pore.

The threshold pressure for gas to invade the corner film at a side wall is $\sigma/r_{c,h}$, larger than those of pores, since $r_{c,h}$ is smaller than pore sizes. To this end, removal of the corner film observed in Fig. 9 could be attributed to the fact that gas invades the corner film at the convex intersection between two side walls (its threshold pressure is smaller than those at side walls), resulting in the interruption of the corner film along the side walls. Such convex intersection at the solid element seems to serve as scissors to cut off the corner films. Therefore, we call this effect as the capillary scissors. Such effect is not considered in the present DPNM_wC. Incorporation of the capillary scissor effect into the DPNM is considered in a future study.

## 6. Conclusions

In summary, a novel dynamic pore network model with corner films (DPNM_wC) is successfully developed to describe evaporation in porous media. The feature of the developed DPNM_wC is inclusion of not only both the continuous and discontinuous corner films as well as the effects of corner films on the threshold pressures of pores but also the liquid viscous forces and the capillary valve effect. The developed DPNM_wC is validated by comparing the simulation results with the evaporation experiment based on a 2D microfluidic pore network of $5 \times 5$ pore bodies. The modeling and experimental results are in good agreement, demonstrating the effectiveness of the developed DPNM_wC.

To further understand the roles of corner films on evaporation in porous media,

the quasi-static pore network model with continuous and discontinuous corner films (QSPNM_wC) and the dynamic pore network model without corner films (DPNM_woC) are also compared with the experiment and the DPNM_wC. The DPNM_woC predicts a lower total evaporation rate, attributed to the fact that the corner films are not considered. Comparison between the QSPNM_wC and the experiment and the DPNM_wC indicates that the liquid viscous forces must be taken into account for accurate description of the evaporation processes in porous media with corner films, even for a pore network with a small size.

Comparison between the DPNM_wC and the experiment further reveals that the continuous corner films, which provide pathways for liquid flow from the inside to the surface of the pore network and contribute to sustain the total evaporation rate, can be interrupted to be the discontinuous ones not only by gas invasion into pores but also by the capillary scissors effect due to the local convex topology of the solid matrix.

Since developing an accurate PNM is the cornerstone to employ it to understand the transport processes in porous media, we believe that the developed DPNM_wC in this study paves a new way for understanding of the multi-scale transport processes during evaporation in porous media with both the continuous and discontinuous corner films.

**Acknowledgement**

The authors are grateful for support of the National Key Research and Development

**Appendix A**

We present in this appendix the detailed calculation procedure to determine the value of $\alpha$ in Eq. (3). Since the inertial and gravity forces are neglected, the liquid flow in corner films is described as:

$$\nabla^2 v_{l,c} = \frac{1}{\mu_l}\frac{dP_{l,c}}{dx} \tag{A1}$$

where $v_{l,c}$ is the liquid velocity in corner films, $\mu_l$ the liquid viscosity, and $x$ the flow direction. Eq. (A1) can be non-dimensionalized as:

$$\nabla^{*2} v_{l,c}^* = 1 \tag{A2a}$$

$$\nabla^* = r_c \nabla \tag{A2b}$$

$$v_{l,c}^* = \frac{\mu_l v_{l,s}}{r_c^2 (\partial P_{l,c}/\partial x)} \tag{A2c}$$

where $r_c$ is the curvature radius of the meniscus in the plane perpendicular to the liquid flow in $x$ direction.

We solve Eq. (A2a) in the cross-sectional zone of the corner film. The lengths of walls at the cross section of the corner film are determined based on the values of $\theta_{si}$, $\theta_g$, and $r_c$. At the wall surfaces, the liquid velocity is zero; at the gas-liquid interfaces, the shear stress is zero. The distributions of the dimensionless liquid velocity in the cross sections of the corner films attached to top and bottom wall are shown in Fig. 11. Based on this liquid velocity distribution, we can get the mass flow rate along the corner film with cross-sectional area $A_c$:

$$Q_c = \int \rho_l v_{l,c}\, da = \int \rho_l \frac{r_c^4}{\mu_l} \frac{dP_{l,c}}{dx} v_{l,c}^*\, da^* \tag{A3a}$$

$$a^* = \frac{a}{r_c^2} \tag{A3b}$$

By comparing Eqs. (3) and (A3a), we can get the dimensionless parameter $\alpha$:

$$\alpha = \int v_{l,c}^*\, da^* \tag{A4}$$

**Appendix B**

We present in this appendix the detailed derivation of Eq. (8). Fig. 12 illustrates schematically a quasi-static gas invasion in a filled pore throat. The arrow shows the invasion direction. The work associated with an infinitesimal advancement of the meniscus towards the liquid phase is equal to the net change in the surface energy:

$$\Delta P A_{gas}\, dx = \sigma_{l,g} dS_{l,g} + \sigma_{silicon,g} dS_{silicon,g} + \sigma_{glss,g} dS_{glss,g} \\ + \sigma_{silicon,l} dS_{silicon,l} + \sigma_{glss,l} dS_{glass,l} \tag{B1}$$

where $dS$ is the change of the area, $\Delta P = P_g - P_l$, and $A_{gas}$ the area of gas zone in the A – A plane shown in Fig. 12.

For gas invasion in a pore throat, we have:

$$dS_{l,g} = l_{l,g} \Delta x \tag{B2a}$$

$$dS_{silicon,g} = -dS_{silicon,l} = l_{silicon,g} \Delta x \tag{B2b}$$

$$\sigma_{silicon,g} - \sigma_{silicon,l} = \sigma_{l,g} \cos\theta_{si} \tag{B2c}$$

$$dS_{glass,g} = -dS_{glass,l} = l_{glass,g} \Delta x \tag{B2d}$$

$$\sigma_{glass,g} - \sigma_{glass,l} = \sigma_{l,g} \cos\theta_g \tag{B2e}$$

where $l_{l,g}$, $l_{silicon,g}$ and $l_{glass,g}$ are the lengths of liquid-gas interface, silicon-gas interface and glass-gas interface shown in A – A plane in Fig. 12, respectively. Substituting Eq.

(B2) into Eq. (B1) yields:

$$\Delta P = \sigma_{l,g} \frac{l_{l,g} + l_{silicon,g}\cos\theta_{si} + l_{glass,g}\cos\theta_g}{A_{gas}} \tag{B3}$$

Based on the topology of the meniscus in the pore throat shown in Fig. 12, we have:

$$l_{l,g} = 2l_{l,g,glass} + 2l_{l,g,silicon} \tag{B4a}$$

$$l_{l,g,glass} = r_c\left(\frac{\pi}{2} - \theta_{si} - \theta_g\right) \tag{B4b}$$

$$l_{l,g,silicon} = r_c\left(\frac{\pi}{2} - 2\theta_{si}\right) \tag{B4c}$$

$$l_{silicon,g} = 2h + w - 4y_1 - 2x_1 \tag{B4d}$$

$$l_{glass,g} = w - 2x_2 \tag{B4e}$$

$$x_1 = r_c - r_c\sin\theta_{si} \tag{B4f}$$

$$x_2 = r_c\cos\theta_{si} \tag{B4g}$$

$$y_1 = r_c\cos\theta_{si} - r_c\sin\theta_{si} \tag{B4h}$$

$$A_{gas} = wh - \left(2A_{silicon,glass} + 2A_{silicon,silicon}\right) \tag{B4i}$$

$$A_{silicon,glass} = r_c^2\cos\theta_{si} - \frac{r_c^2}{2}\cos\theta_{si}\sin\theta_{si} - \frac{\pi - 2\theta_{si}}{4}r_c^2 \tag{B4j}$$

$$A_{silicon,silicon} = r_c^2\cos^2\theta_{si} - r_c^2\sin\theta_{si}\cos\theta_{si} - \frac{\pi - 4\theta_{si}}{4}r_c^2 \tag{B4k}$$

Substituting Eq. (B4) into Eq. (B3) yields:

$$\Delta P = \sigma_{l,g}\frac{k_1 + 2k_2 r_c}{wh - 2k_3 r_c^2} \tag{B5a}$$

$$k_1 = (2h + w)\cos\theta_{si} + w \tag{B5b}$$

$$k_2 = \pi - 3\theta_{si} - (2\cos\theta_{si} - 3\sin\theta_{si} + 2)\cos\theta_{si} \tag{B5c}$$

$$k_3 = \cos^2\theta_{si} + \cos\theta_{si} - \frac{3}{2}\sin\theta_{si}\cos\theta_{si} - \frac{\pi}{2} + \frac{3\theta_{si}}{2} \tag{B5d}$$

The value of $\Delta P$ depends on $r_c$. The minimum $\Delta P$ is the threshold pressure for gas to invade the pore, which can be calculated by setting the derivative of $\Delta P$ with respect

to $r_c$ to be zero:

$$\frac{dP}{dr_c} = \sigma_{l,g}\left[\frac{2k_2}{wh - 2k_3 r_c^2} + \frac{4k_3 r_c(k_1 + 2k_2 r_c)}{(wh - 2k_3 r_c^2)^2}\right] = 0 \tag{B6}$$

By solving Eq. (B6), the value of $r_c$ at the minimum $\Delta P$ is expressed as:

$$r_c^* = \sqrt{\frac{k_1^2}{4k_2^2} - \frac{wh}{2k_3}} + \frac{k_1}{2k_2} \tag{B7}$$

To this end, the threshold pressure to invade a pore throat of width $w$ is:

$$P_t = -\sigma \frac{k_1 + k_2 r_c^*}{wh - k_3 r_c^{*2}} \tag{B8}$$

The values of $k_1$, $k_2$, $k_3$, and $r_c^*$ can be found in Eqs. (B5) and (B7).

**Figure Captions**

**Fig. 1** Schematic of continuous and discontinuous corner films in porous media.

**Fig. 2** (a) Schematic of the microfluidic pore network used in the experiment. (b) Cross section of the outlet pore in the pore network. (c) Schematic of corner films in a pore.

**Fig. 3** Swelling of continuous corner films and shrinkage of liquid during evaporation in the pore network.

**Fig. 4** Variation of liquid distribution in the pore network during evaporation obtained from experiment, DPNM_wC, and DPNM_woC. $S_t$ is the total liquid saturation in the pore network.

**Fig. 5** Variation of the number of liquid clusters during evaporation in the DPNM_wC and the DPNM_woC.

**Fig. 6** Gas invasion processes in the upper-right zone of the pore network during evaporation in the (a) DPNM_wC and (b) DPNM_woC.

**Fig. 7** Variation of the total liquid saturation, $S_t$, with the evaporation time, $t$.

**Fig. 8** Variation of the liquid velocity in pore throat B shown in Fig. 4 during evaporation.

**Fig. 9** Gas invasion process in upper-left zone of the pore network during evaporation.

**Fig. 10** Schematic of the morphology of the meniscus at the intersection between two side walls.

**Fig. 11** Distribution of the dimensionless liquid velocity in corner films.

**Fig. 12** Schematic of gas invasion in a pore throat. (a) Top view. (b) Side view. (c) Distribution of corner films in the A –A plane. Liquid is shown in red, and gas is shown in white. The arrow shows the gas invasion direction.

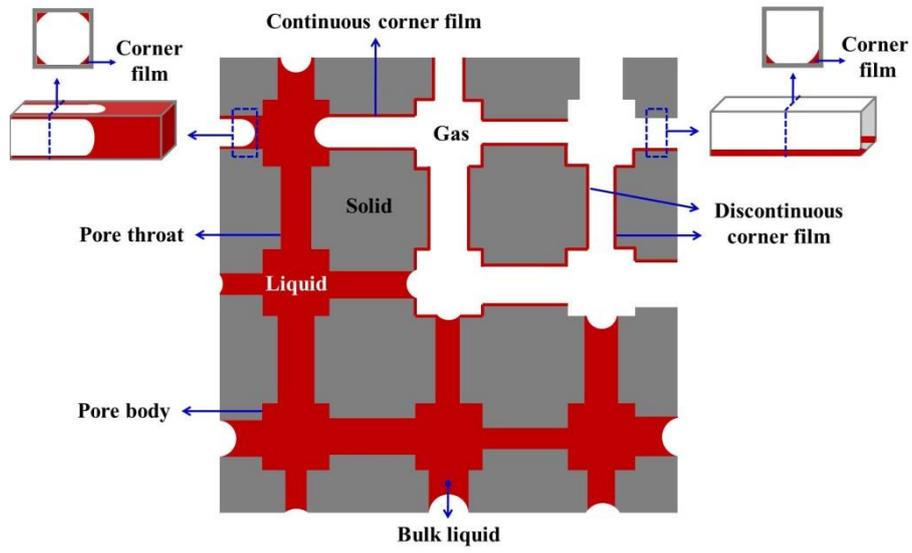

Fig. 1

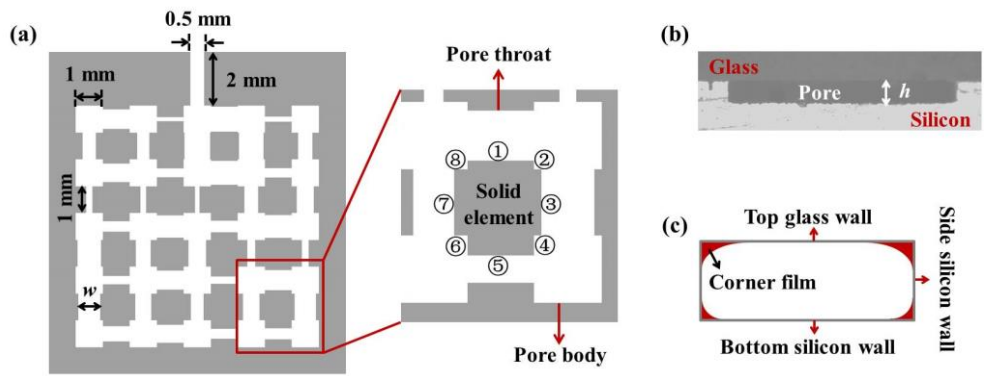

Fig. 2

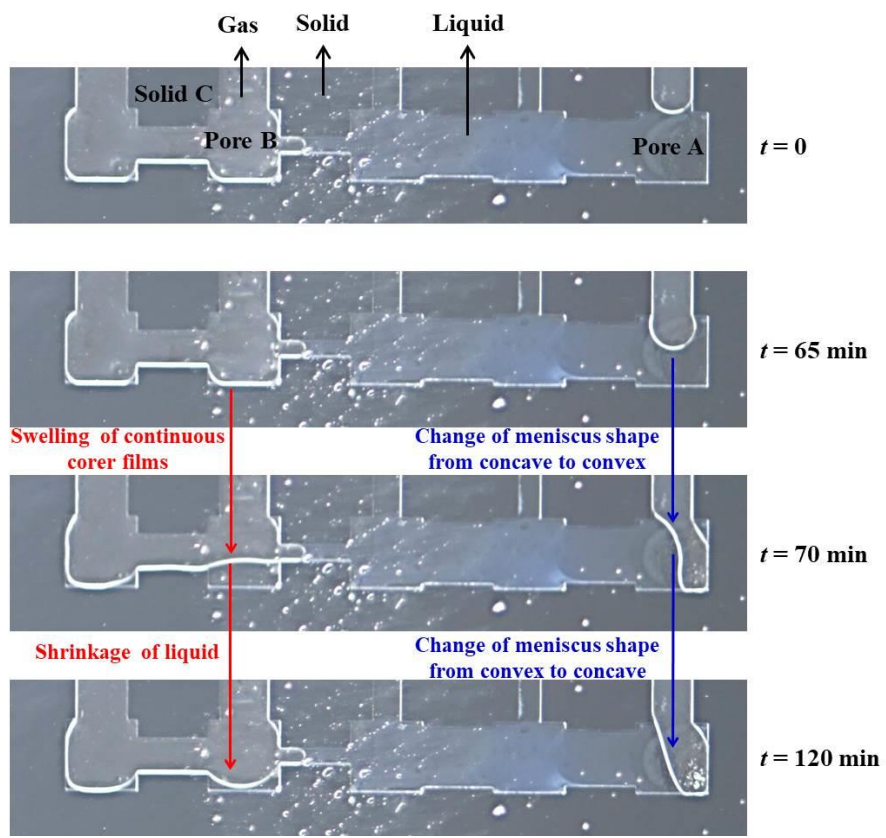

Fig. 3

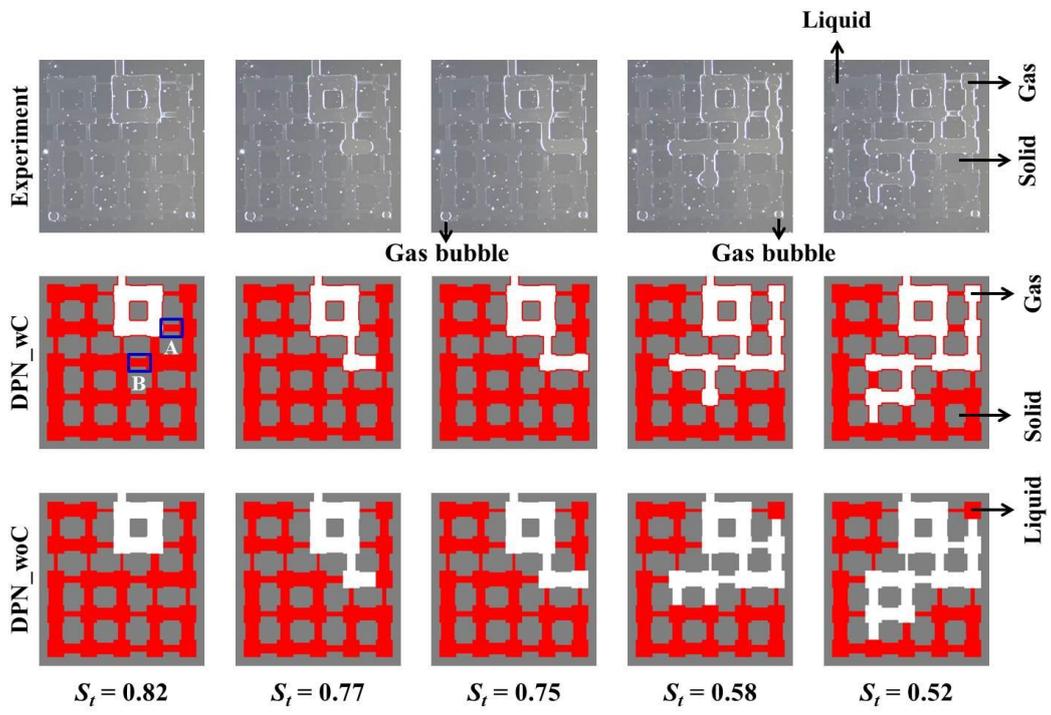

Fig. 4

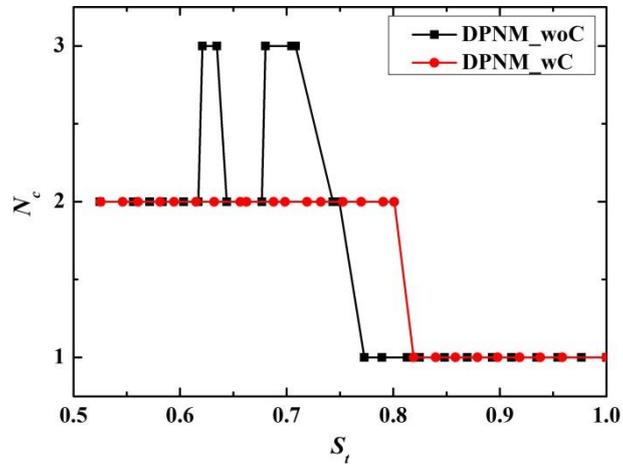

Fig. 5

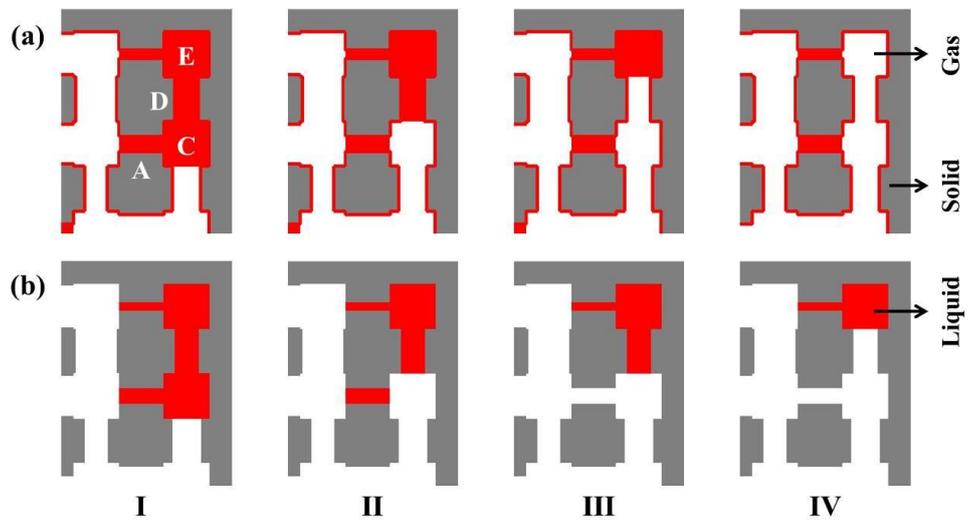

Fig. 6

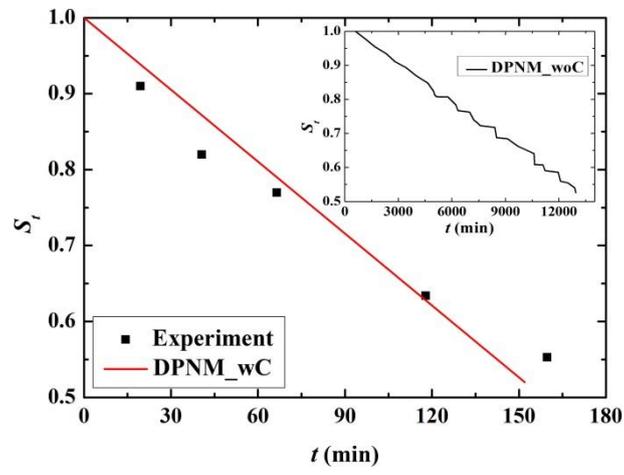

Fig. 7

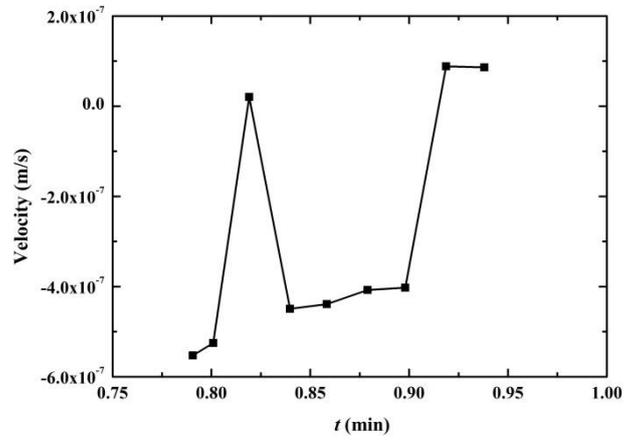

Fig. 8

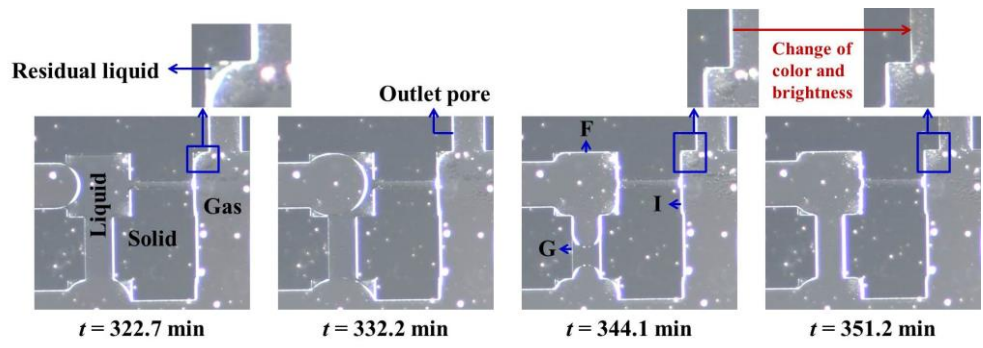

Fig. 9

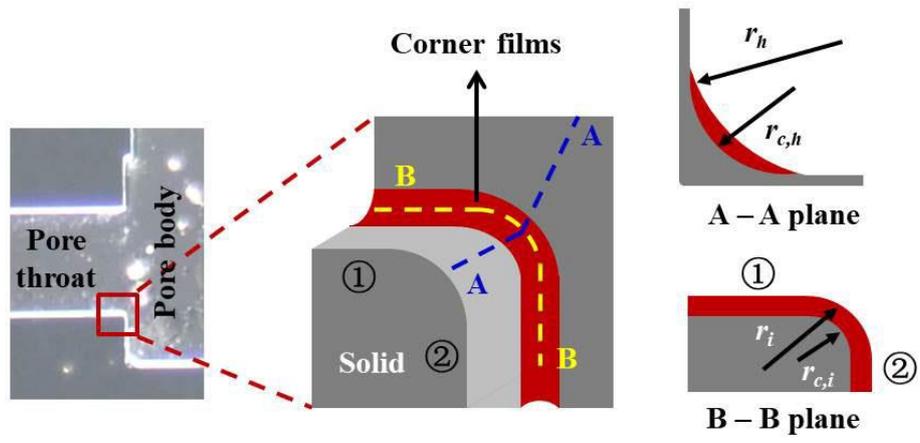

Fig. 10

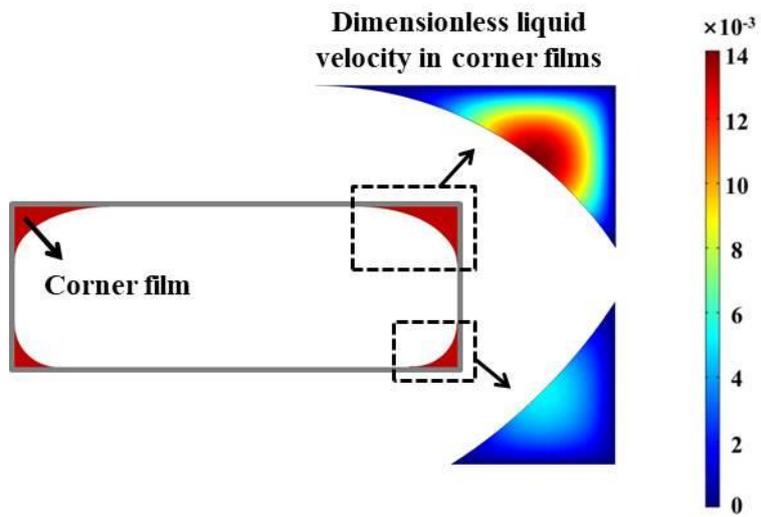

Fig. 11

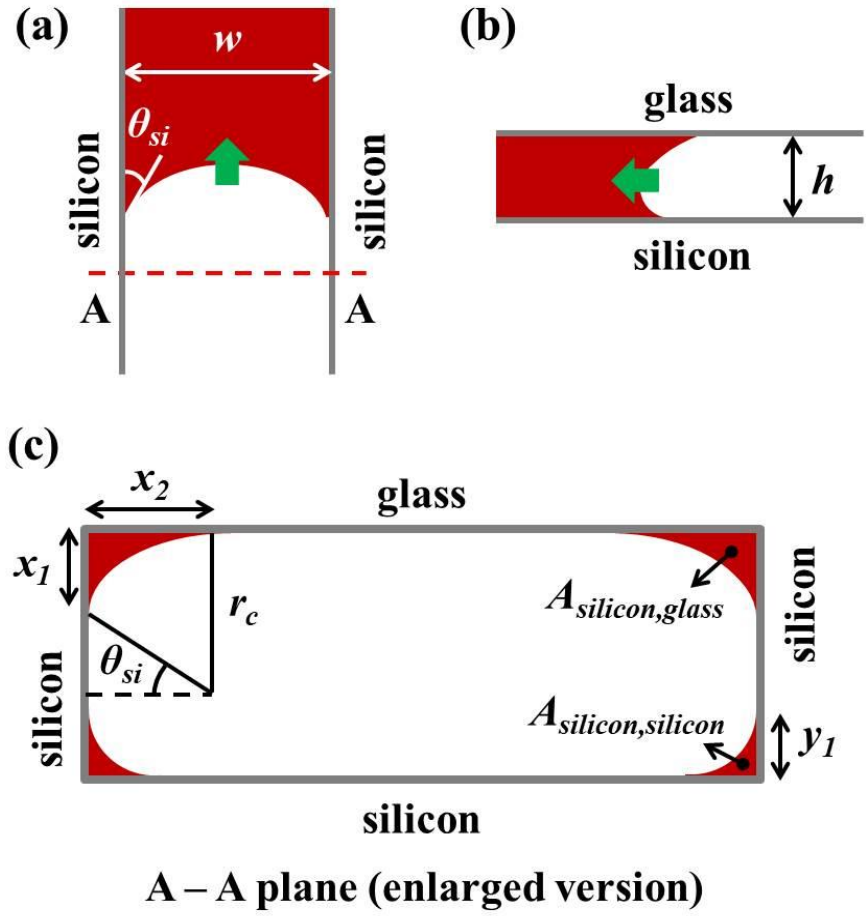

Fig. 12

# Pore network model of evaporation in porous media with continuous and discontinuous corner films: Supporting materials

Rui Wu[a,b], Tao Zhang[a], Chao Ye[a], C.Y. Zhao[a,b,*], Evangelos Tsotsas[c], Abdolreza Kharaghani[c]

[a]School of Mechanical Engineering, Shanghai Jiao Tong University, Shanghai 200240, China.

[b]Key Laboratory for Power Machinery and Engineering, Ministry of Education, Shanghai Jiao Tong University, Shanghai 200240, China

[c]Chair of Thermal Process Engineering, Otto von Guericke University, P.O. 4120, 39106 Magdeburg, Germany

* Corresponding author Tel: +86 (0)21-34204541

E-mails:

ruiwu@sjtu.edu.cn (R. Wu)

changying.zhao@sjtu.edu.cn (C.Y. Zhao).

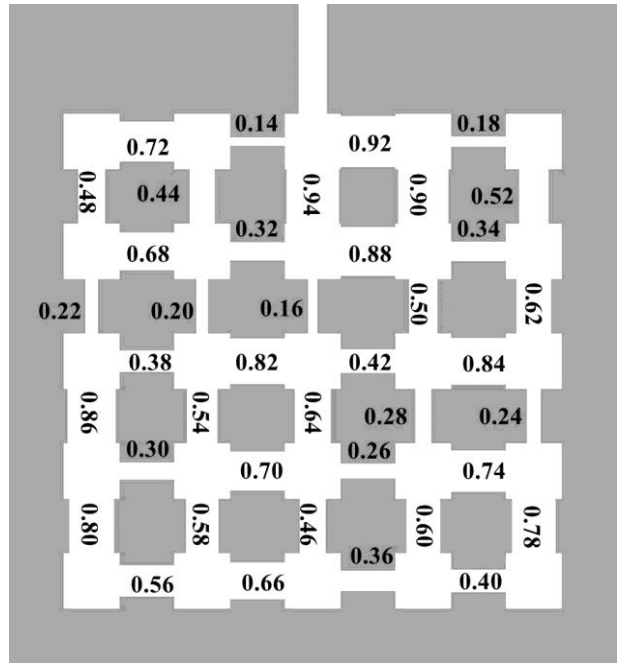

Fig. S1 Size of each pore throat of the pore network. The number is the width of the pore throat; the unit is mm.

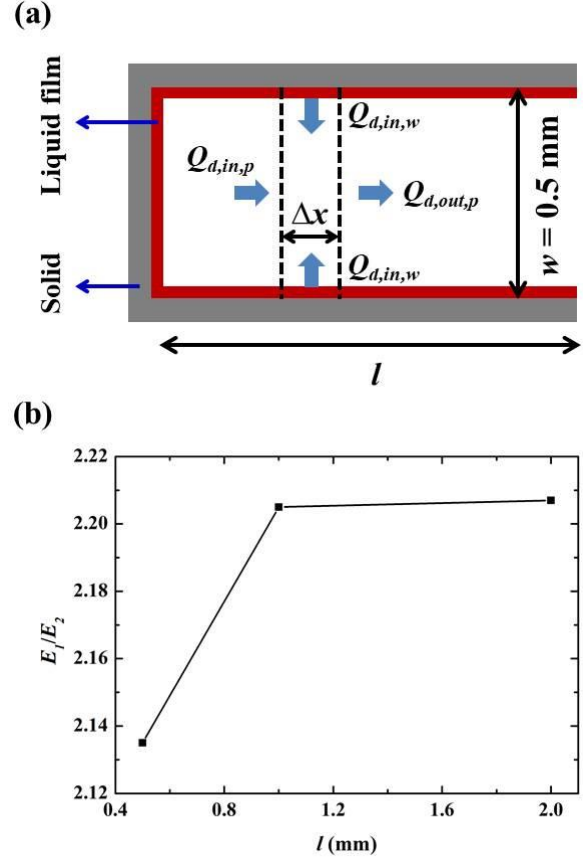

Fig. S2 (a) Schematic of 1D + 1D model for vapor transport in a pore with saturated side walls. $Q_{d,in,p}$ and $Q_{d,in,w}$ are the rate of vapor diffusion into the control volume from the adjacent control volume and side wall, respectively. $Q_{d,out,p}$ is the rate of vapor diffusion out of the control volume. Based on mass conservation law, $Q_{d,out,p} = Q_{d,in,p} + Q_{d,in,w}$. Based on this equation, we can get $\frac{d^2 \ln(P_g - P_v)}{dx^2} = 4 \frac{\ln(P_g - P_v) - \ln(P_g - P_s)}{w}$. The pore width is $w = 0.5$ mm. The vapor pressure at the wall surface is $P_v = P_s$, for which $P_s$ is the saturated vapor pressure. At the entrance of the pore, $P_v = 0$. (b) Ratio of the evaporation rate from the pore obtained from 2D model, $E_1$, to that obtained from 1D +1D model, $E_2$. This ratio is about 2, indicating that 1D + 1D cannot describe accurately the vapor transport in the pore.

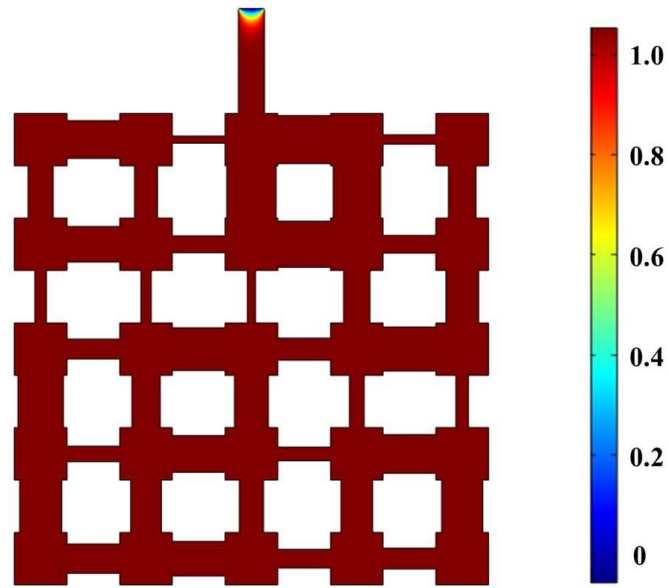

Fig. S3 Vapor pressure field (normalized to the saturated vapor pressure) in the pore network when the side walls at edges of the pore network and at outlet pore are saturated, while the other side walls are empty. We can see that the vapor in the pore network is saturated.

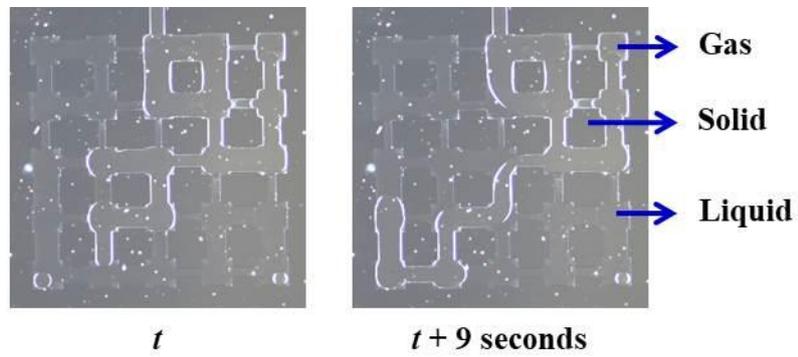

Fig. S4 Meniscus instability observed in the experiment. As gas invades the pore body at the bottom of the pore network, the meniscus in this pore body becomes unstable, and quickly invades the pores at the lower-left zone of the pore network (in about 9 seconds); at the same time, some pore near the center of the pore network are refilled by liquid.

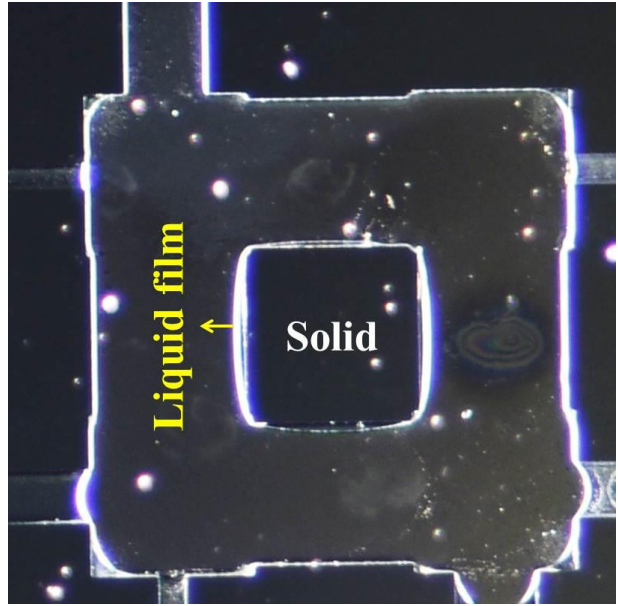

Fig. S5 Experimentally observed liquid films around the solid element near the outlet pore.

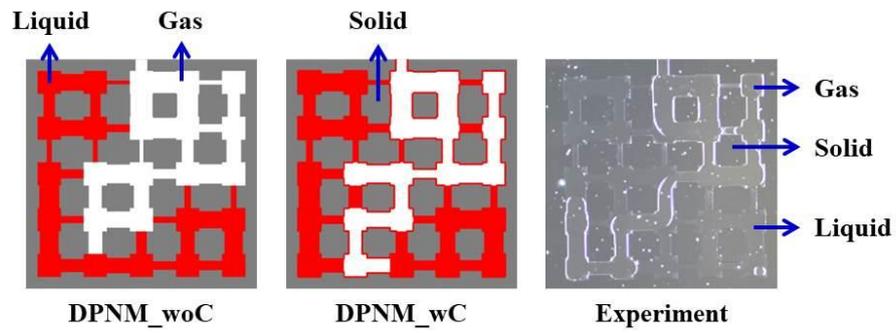

Fig. S6 Liquid distribution in the pore network at $S_t = 0.49$. Left: simulation predicted by DPNM_woC. Middle: simulation predicted by DPNM_wC. Right: experiment.

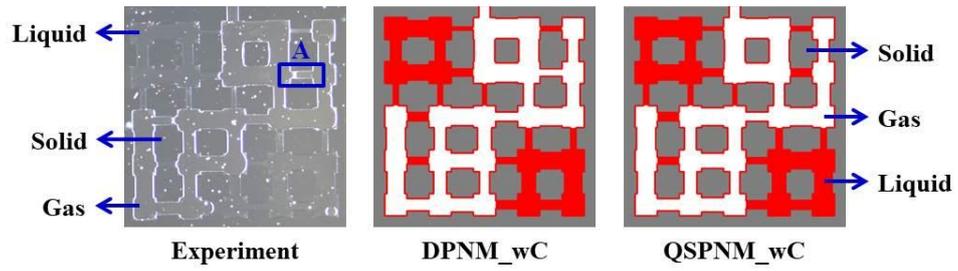

Fig. S7 Liquid distribution in the pore network when pore throat A is empty. Left: experimental results. Middle: DPNM_wC. Right: simulation results from QSPNM_wC.

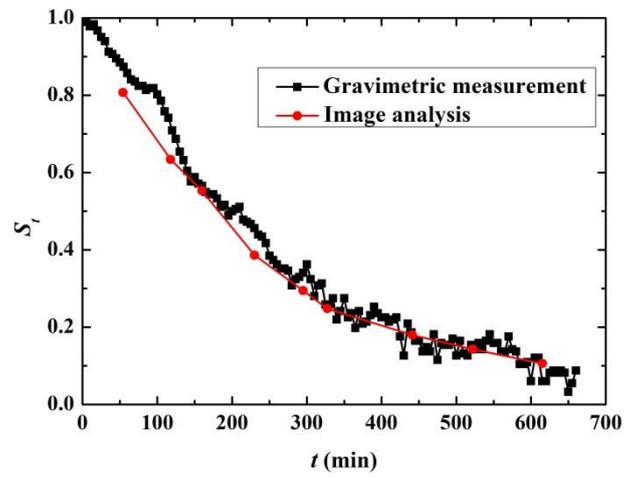

Fig. S8 Evolution of the total liquid saturation, $S_t$, with the evaporation time, $t$, obtained from two experimental methods: image analysis and gravimetric measurement.

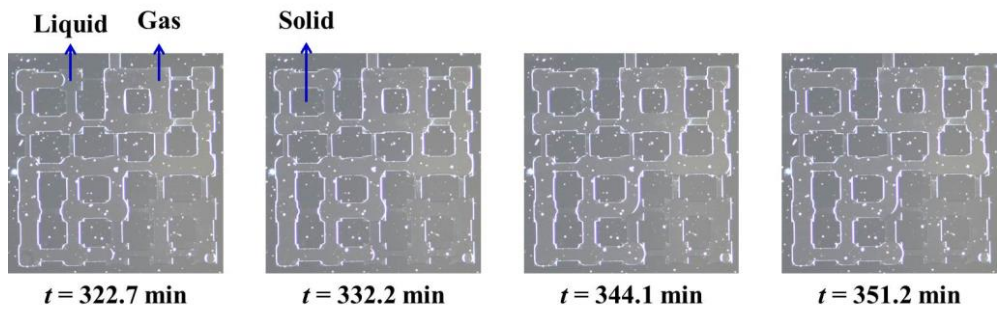

Fig. S9 Evolution of liquid distribution in the pore network during evaporation ($t$ = 322.7 min and $t$ = 351.2 min correspond to $S_t$ = 0.25 and $S_t$ = 0.18, respectively).

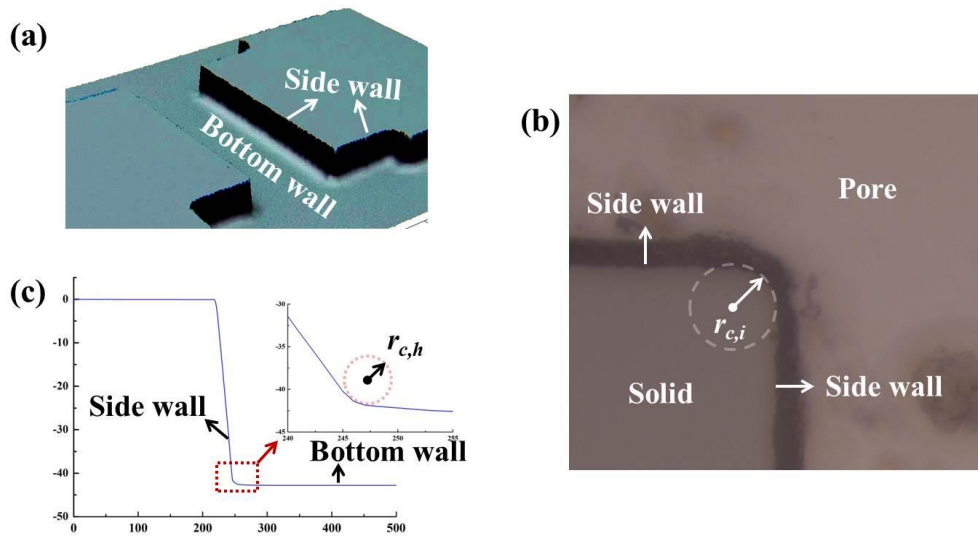

Fig. S10 (a) 3D structure of the pore in the pore network obtained by a 3D microscope (Zeta 200). (b) Intersection between two side walls; this intersection is rounded. (c) Profile of a side wall gained by a surface profiler (Ambios XP-200); the corner between the side wall and bottom wall is not sharp but rounded.